\documentclass[aps,prl,twocolumn,amsmath,amssymb,superscriptaddress]{revtex4-2}

\usepackage{upgreek}
\usepackage{graphicx} 
\usepackage{bm} 
\renewcommand{\figurename}{Figure}
\renewcommand{\tablename}{Table}
\makeatletter
\def\fnum@figure{\textbf{\figurename~\thefigure}}
\def\fnum@table{\textbf{\tablename~\thetable}}
\makeatother

\renewcommand{\vec}[1]{\bm{\mathbf{#1}}}

\newcommand{\He}[1]{\textsuperscript{#1}He}

\usepackage{color}

\begin{document}

\title{Cooling Low-Dimensional Electron Systems into the Microkelvin Regime}

\author{Lev\ V.\ Levitin}
\email{l.v.levitin@rhul.ac.uk}
\affiliation{Department of Physics, Royal Holloway, University of London, Egham TW20 0EX, UK}
\author{Harriet\ van der Vliet}
\altaffiliation[Now at ]{Oxford Instruments Nanoscience, Abingdon, Oxfordshire OX13 5QX, UK.}
\affiliation{Department of Physics, Royal Holloway, University of London, Egham TW20 0EX, UK}
\author{Terje\ Theisen}
\affiliation{Department of Physics, Royal Holloway, University of London, Egham TW20 0EX, UK}
\author{Stefanos\ Dimitriadis}
\altaffiliation[Now at ]{Department of Physics, Imperial College London, London SW7 2AZ, UK.}
\affiliation{Department of Physics, Royal Holloway, University of London, Egham TW20 0EX, UK}
\author{Marijn~Lucas}
\affiliation{Department of Physics, Royal Holloway, University of London, Egham TW20 0EX, UK}
\author{Antonio\ D.\ Corcoles}
\altaffiliation[Now at ]{Thomas J. Watson Research Center, Yorktown Heights, NY 10598, USA.}
\affiliation{Department of Physics, Royal Holloway, University of London, Egham TW20 0EX, UK}
\author{J\'{a}n Ny\'{e}ki}
\affiliation{Department of Physics, Royal Holloway, University of London, Egham TW20 0EX, UK}
\author{Andrew\ J.\ Casey}
\affiliation{Department of Physics, Royal Holloway, University of London, Egham TW20 0EX, UK}
\author{Graham~Creeth}
\altaffiliation[Now at ]{Praesto Consulting, Dublin, D02 A342, Ireland.}
\affiliation{London Centre for Nanotechnology, University College London, London WC1H 0AH, UK}
\author{Ian\ Farrer}
\altaffiliation[Now at ]{Department of Electronic and Electrical Engineering, University of Sheffield, Sheffield S1 3JD, UK.}
\affiliation{\mbox{Cavendish Laboratory, University of Cambridge, JJ Thomson Avenue, Cambridge CB3 0HE, UK}}
\author{David\ A.\ Ritchie}
\affiliation{\mbox{Cavendish Laboratory, University of Cambridge, JJ Thomson Avenue, Cambridge CB3 0HE, UK}}
\author{James\ T.\ Nicholls}
\affiliation{Department of Physics, Royal Holloway, University of London, Egham TW20 0EX, UK}
\author{John\ Saunders}
\affiliation{Department of Physics, Royal Holloway, University of London, Egham TW20 0EX, UK}

\date{January 26, 2022}

\begin{abstract}
Two-dimensional electron gases (2DEGs) with high mobility, 
engineered in semiconductor heterostructures
host a variety of ordered phases arising from strong correlations, which emerge at sufficiently low temperatures.
The 2DEG can be further controlled by surface gates to create quasi-one dimensional systems, with potential spintronic applications. 
Here we address the long-standing challenge of cooling such electrons to below 1\,mK, potentially important
for identification of topological phases and spin correlated states. 
The 2DEG device was immersed in liquid \He3, cooled by the nuclear adiabatic demagnetization of copper. 
The temperature of the 2D electrons was inferred from the electronic noise in a gold wire,
connected to the 2DEG by a metallic ohmic contact. With effective screening and filtering,
we demonstrate a temperature of 0.9$\,\pm\,$0.1\,mK, with scope for significant further improvement. 
This platform is a key technological step, paving the way to observing new quantum phenomena,
and developing new generations of nanoelectronic devices exploiting correlated electron states.
\end{abstract}

\maketitle

\section{Introduction}
Two-dimensional electron gases (2DEGs), created at a GaAs-AlGaAs heterojunction and grown by molecular-beam epitaxy (MBE),
have been the building block for the study of low-dimensional physics over the past few decades. 
This requires the high crystal perfection of MBE growth and the use of modulation doping (see Ref.~\cite{Umansky2009} for a review
of the state of the art).

The lowest possible electron carrier temperature coupled to the highest sample quality is key to discovering and elucidating
new ground states in such systems. The observation of the integer quantum Hall effect~\cite{Klitzing81} and shortly thereafter
the fractional quantum Hall effect (FQHE)~\cite{Tsui1982} has led to extensive studies of two-dimensional electron systems
in perpendicular magnetic field $B$. The lowest Landau level (LL), with filling factors $\nu<1$ (here $\nu=n h /eB$ is determined
by the carrier density $n_{\text{2D}}$, Planck's constant $h$ and elementary charge $e$)
for the lowest spin branch is described by composite fermion (CF) theory~\cite{Jain1989}.   
It exhibits a rich variety of quantum states:  fractional quantum Hall states for odd fractions,
while $\nu=1/2$ is a composite fermion Fermi sea~\cite{Kamburov2014}.
A key question is the role of residual CF interactions in giving rise to exotic states~\cite{Jain2015}. 
A Wigner solid is found for small $\nu$, which also exhibits re-entrance with a maximum melting temperature of 50\,mK
above the $\nu=1/3$ FQHE state~\cite{Ma2020}. 
Interactions are also believed to play a role at other fractional fillings such as $\nu= 4/11$,
which has a measured activation energy of 5\,mK~\cite{Pan2015}.

\begin{figure}[t!]
\centerline{\includegraphics{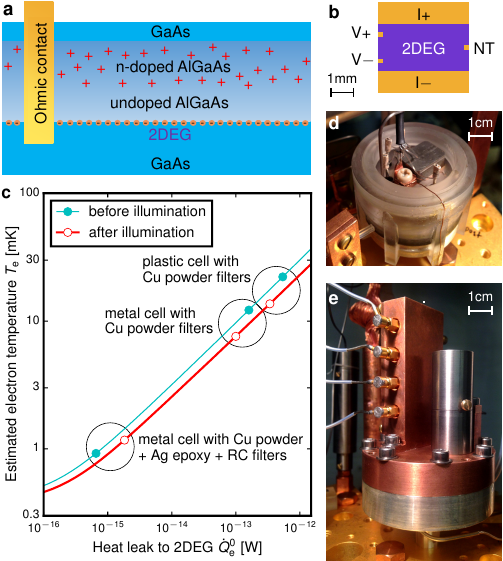}}
\caption{\textbf{Cooling electrons in a 2DEG in \He3 immersion cells.}
\textbf{a}~Schematic of the sample: the 2DEG is created at a GaAs/AlGaAs heterojunction,
and electrical connections are provided by \mbox{AuNiGe} ohmic contacts.
\textbf{b}~Top view of the 4\,mm$\times$4\,mm device cooled to 1\,mK by the I$\pm$ and V$\pm$ ohmic contacts,
also used to measure the electrical resistances in the device.
A noise thermometer (NT) is connected to the NT ohmic contact to probe the electron temperature $T_{\text e}$.
\textbf{c}~Progressive cooling of the 2DEG device in plastic (\textbf{d}) and metal (\textbf{e}) immersion cells
equipped with different low-pass filters.
$T_{\text e}$ is estimated from Eq.~\eqref{eq:dotQ:Te},
assuming that the \He3 bath is at temperature $T_{\text{bath}} = 0.3$\,mK.}
\label{fig:immcells}
\end{figure}

The nature of quantum states in the second LL is in many cases not resolved, and this 2D electron system is of particular interest
for study to lower temperatures than hitherto achieved. We discuss a few examples here. Even denominator states forming in the second Landau level
are thought to arise from pairing of composite fermions. Confirmation of exact quantization of the 5/2 state
first discovered by Willett {\em et al.}~\cite{Willett1987} required electron temperatures as low as 8\,mK~\cite{Pan1999}. 
Nevertheless the topology of the $\nu=5/2$ ground state (Pfaffian, anti-Pfaffian or other) remains an open question. 
Observations are highly sensitive to disorder~\cite{Mross2018,Wang2018}.
It has been shown that a quantum phase transition to a quantum Hall nematic phase can be induced by applied pressure
in both the 5/2 and 7/2 state~\cite{schreiber2018}. 
The importance of such topological ground states is reinforced by the proposal that they offer a route to topological quantum computing
using non-Abelian anyons~\cite{Nayak2008}. A non-Abelian phase is also predicted for the 12/5 state.

Many of these states have small energy gaps. The 12/5 state has a gap of order 30\,mK, which exhibits non-trivial tilt dependence of the magnetic field.
Studies of the energy gaps at other fractional fillings~\cite{Kumar2010,Shingla2018} found that the gaps of the $\nu= 2+6/13$ (10\,mK)
and $2+2/5$ (80\,mK) states do not follow the CF hierarchy, leading to the proposal of a parton state~\cite{Balram2018}.
The $2+3/8$ state also has a reported gap of only 10\,mK. Furthermore there are indications of novel topological order in the upper
spin branch of the second LL, at fillings $3+1/3$ (37\,mK) and $3+1/5$ (104\,mK)~\cite{Kleinbaum2015}.

New states arising from strong correlations in the low-dimensional electron system are also expected in low magnetic fields.
Electrostatic confinement of the 2DEG to form 1D wires and zero-dimensional (0D) quantum dots has led to studies of the
many-body physics of Luttinger liquids~\cite{Auslaender2002,Jompol2009} and the Kondo effect~\cite{Potok2007, Iftikhar2015}.
Furthermore the conduction electrons in GaAs-based 2DEGs are expected to couple to the nuclear spins, undergoing a collective
ferromagnetic transition mediated by RKKY interactions which is expected at electron temperatures in the mK regime~\cite{Simon2008}.
The hyperfine interaction is also predicted~\cite{Braunecker2013} to create new topological phases in a semiconductor 1D wire coupled
to a superconductor and magnetic moments.
A transition from metal- to insulator-like behaviour with decreasing temperature has been reported in a 2D hole gas in GaAs~\cite{Huang2007},
potentially related to many-body localization. Recently, Wigner crystal-like zigzag chains have been discovered in 1D wires
at low electron densities~\cite{Ho2018}. At lower temperatures there is potential for of quantum communication through 1D spin chains,
coupling high fidelity quantum dot qubits~\cite{Bose2007}.

Advances in cryogen-free technologies now make low millikelvin temperature platforms widely accessible~\cite{Batey2009},
with prospects of commercial solutions in the microkelvin range~\cite{Batey2013,Todoshenko2014}.
However cooling electrons in semiconductor devices is challenging.
In situ cooling via the electron-phonon coupling is severely limited by its strong temperature dependence,
while the electrons are susceptible to heat leaks arising from electromagnetic noise.
Electromagnetic shielding and ultra-sensitive measurement schemes with low or no electrical 
excitation~\cite{Hashisaka2009,Scheller2014,Iftikhar2016} are common strategies for minimising the heat leak.
Cooling via the electrical leads has been widely implemented~\cite{Iftikhar2016,Samkharadze2011,Pan1999,Xia2000,Nicoli2019},
commonly employing immersion in liquid helium. 
For 2D electrons, this has so far been limited, for many years, to 4\,mK, see Ref.~\cite{Samkharadze2011} and references therein.
Primary thermometers within samples confirmed 2D electron temperatures of 6\,mK~\cite{Iftikhar2016,Nicoli2019}.

\begin{figure*}[t!]
\centerline{\includegraphics{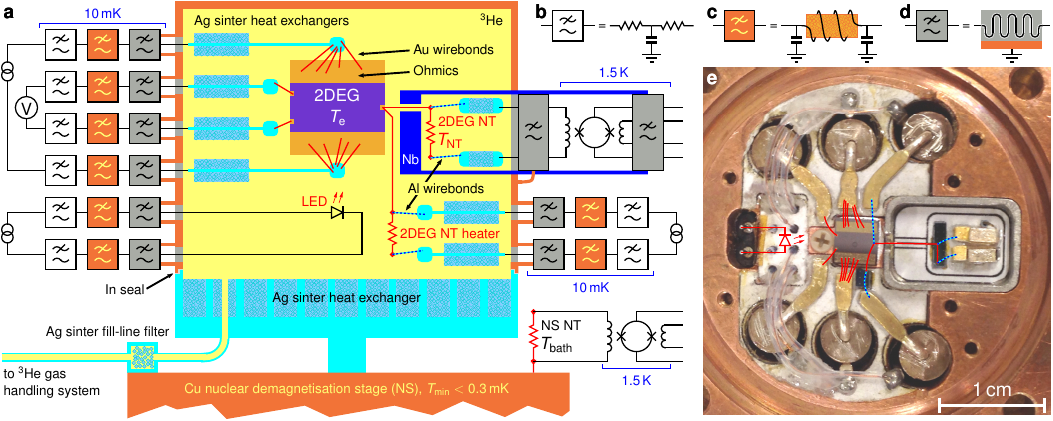}}
\caption{\textbf{Experimental platform for cooling low-dimensional electrons.}
\textbf{a}~Schematic diagram of the metal immersion cell.
Silver sinter on the silver base thermalises the liquid \He3 to the nuclear
demagnetisation refrigerator. The copper lid holds the 2DEG device, the sinter heat exchangers
for cooling the I$\pm$ and V$\pm$ lines, the 2DEG noise thermometer (NT), and a red light-emitting diode (LED) for illuminating the 2DEG.
The NT and its heater were connected using wedge-bonded resistive Au wires and
superconducting Al wires acting as thermal breaks~\cite{Ventura1998}.
Another NT is mounted on the nuclear stage (NS) to measure the \He3 bath temperature $T_{\text{bath}}$.
The base and the lid sealed with In comprise a He-tight and photon-tight enclosure, complete with a Ag sinter plug in the filling capillary.
All incoming electrical connections are fitted with low-pass filters of three types:
\textbf{b} discrete $RC$ filters with $R=1$\,k$\Omega$ and $C = 10$\,nF,
\textbf{c} insulating Cu powder filters~\cite{Lukashenko2008}, and,
\textbf{d} Ag epoxy filters with 1\,m-long NbTi wires embedded in conducting Ag epoxy.
RF-tight coaxial connections were used between these filters and the walls of the immersion cell.
Ag epoxy filters with multiple wires, compatible with SQUID NT readout, were used on the SQUID lines.
\textbf{e} The inside of the copper lid of the cell, with the Au and Al wires highlighted and the LED shown schematically.
The NT is housed in a Nb shield, shown here with the top removed.}
\label{fig:metalcell}
\end{figure*}

In response to the challenge to cool nano-electronic devices to ultra-low temperatures (ULT), new approaches have been introduced
which focus on refrigeration by nuclear demagnetisation of each lead connected to the sample~\cite{Clark2010}.
More recently this has been extended by incorporating nuclear refrigeration elements into the device itself.
For this on-chip cooling the Coulomb blockade device used is itself a thermometer,
providing an accurate measurement of the electron temperature in mesoscopic metallic islands as low as 0.5\,mK~\cite{Jones2020,Sarsby2020}.

In contrast, and complementary to this approach, the focus of our work is the cooling of a relatively large area (8\,mm$^2$) 2DEG.
Our approach to cooling is motivated by the requirement for flexibility to cool a wide range of devices in different sample environments.
We report an ultimate electron temperature in the 2DEG of $0.9\pm 0.1$\,mK, cooled in a \He3 immersion cell,
achieved after improvements in electromagnetic shielding. The 2DEG device is immersed in liquid \He3 and cooled through metallic ohmic contacts,
which are coupled to the \He3 via compact heat exchangers made of sintered Ag powder.
Cooling of devices by the immersion cell technique, as compared to on-chip cooling, is remarkably versatile due to modularity of the design. 
The cooling is provided by a physically remote copper nuclear adiabatic demagnetization stage module, 
so the magnetic field applied to the sample can be independently adjusted. A broad class of devices can potentially be cooled in this way, 
and in some cases the \He3 liquid pressure can be used as an experimental control parameter~\cite{Lane2020}.

The effective cooling of electrons in the 2DEG relies on both reducing the resistance (and hence thermal resistance via the Wiedemann-Franz law)
of the 2DEG and ohmic contacts, and reducing the heat leak to the device. We fabricated a high-mobility GaAs-based 2DEG with sub-$1\,\Omega$
AuNiGe ohmic contacts, see Fig.~\ref{fig:immcells}a,b, and reached a 1~fW level of heat leak to the 2DEG, see Fig.~\ref{fig:immcells}c, through designing
a metal cell that is tight to photons over a wide frequency range, combined with extensive filtering of all electrical lines, see Fig.~\ref{fig:metalcell}.

A key feature of the experiment was the choice of thermometer. We use a current sensing noise thermometer (NT) as an external module
attached to the 2DEG via an ohmic contact. A SQUID current sensor is used to read out the voltage fluctuations across a gold wire~\cite{Casey2014}.
Due to the fundamental Nyquist relation the NT operates over 5 orders of magnitude in temperature with a single point calibration.
It can be configured as a primary thermometer~\cite{Shibahara2016}, and as such underpins the recent redefinition of the Kelvin.

The noise thermometer dissipates no power in the device, but we must take account of both the inevitable residual heat leak
into the thermometer, and the potential parallel cooling channel by direct coupling to the \He3 in which it is immersed.
We have constructed a thermal model which demonstrates that at the lowest temperatures the NT accurately reflects the electron temperature in the 2DEG. 
The ability to alter the resistance of the 2DEG in situ by optical illumination, 
due to persistent photoconductivity, was used to experimentally validate the model.

\section{Results}

\begin{figure*}[t!]
\centerline{\includegraphics{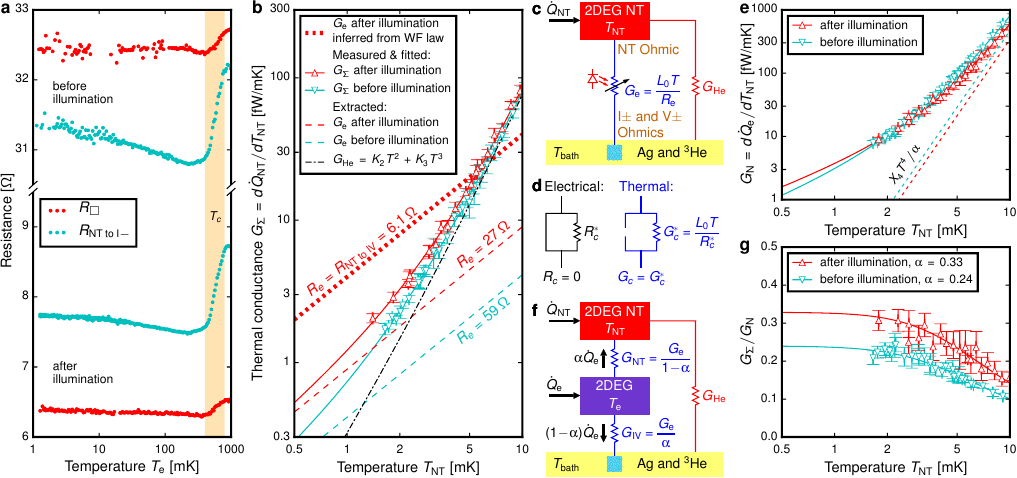}}
\caption{\textbf{Transport measurements of 2DEG and ohmic contacts.}
\textbf{a}~Sheet resistance $R_{\Box}$ of the 2DEG (red)
and the resistance including 2DEG and two ohmic contacts (cyan) measured down to 1\,mK.
Weak temperature dependence of the resistance is observed 
in both the 2DEG and the ohmic contacts below the superconducting transition in the contacts which occurs at $T_c = 0.6 \pm 0.2$~K.
The electron temperature $T_{\text e}$ is determined from Eq.~\eqref{eq:dotQ:Te}.
\textbf{b}~Thermal conductance $G_\Sigma$ between 2DEG NT and \He3 bath before (cyan) and after (red) illuminating the 2DEG,
modelled as a sum of two parallel channels (\textbf{c}): $G_{\text e}$ via electrons in the 2DEG and ohmic contacts, sensitive to illumination,
and $G_{\text{He}}$ due to immersion of the gold wires in liquid helium. Assuming the Wiedemann-Franz (WF) law applies in the 2DEG and ohmic contacts, 
after illumination the electron channel $G_{\text e}^{\text{WF}} = L_0 T / R_{\text{NT\,to\,IV}}$ 
alone exceeds the measured $G_\Sigma$. We attribute the violation of the WF law to superconductivity within the ohmic contacts.
The effective electrical resistance $R_{\text e}$ is obtained by fitting $G_\Sigma(T)$ in \textbf{b} to Eq.~\eqref{eq:G}.
\textbf{d}~In the partially-superconducting ohmic contact model the thermal conductance at $T\ll T_c$ is determined
by the non-superconducting channel $R_c^*$ according to the WF law.
\textbf{e}~Response of the 2DEG NT to heating the 2DEG, the non-local thermal conductance $G_{\text{N}}$, 
is obtained analogously to $G_\Sigma$ and fitted to Eq.~\eqref{eq:GN}.
Dashed lines show the $X_4 T^4$ term in Eq.~\eqref{eq:GN}, unimportant below $T_{\text{NT}} = 3$\,mK.
\textbf{f}~A lumped-element thermal model described by Eqs.~\eqref{eq:dotQ:Tnt} and \eqref{eq:dotQ:Te},
in which $\dot Q_{\text e}$ is assumed to be applied in the centre of the 2DEG,
splitting $G_{\text e}$ into $G_{\text{NT}}$ and $G_{\text{IV}}$, each including the 2DEG and the ohmic contacts.
\textbf{g}~The ratio $G_\Sigma / G_{\text{N}}$ describes the fraction of $\dot Q_{\text e}$ that flows into the NT. 
The $T\to 0$ limit $\alpha$ of $G_\Sigma / G_{\text{N}}$ governs $G_{\text{NT}}$ and $G_{\text{IV}}$ in \textbf{f}.
Error bars in \textbf{b}, \textbf{e} and \textbf{g} represent s.d.}\label{fig:G}
\end{figure*}

\textbf{Experimental setup.}
Figure~\ref{fig:metalcell}a is a schematic of the metal immersion cell,  which combines a \He3 bath with a photon-tight environment.
Filtered connections to room-temperature electronics are provided to probe electrical resistances in the device and to operate heaters
in thermal transport experiments. Two NTs are connected to the 2DEG and the demagnetisation refrigerator.
Illumination by a red light-emitting diode (LED) facing the 2DEG allows us to increase the 2D carrier density and mobility~\cite{Klem1983}.
Preliminary studies were conducted in a plastic cell shown in Fig.~\ref{fig:immcells}d, which was based on Ref.~\cite{Samkharadze2011}.
See Supplementary Information (SI) for further details.

\textbf{Electrical transport measurements.}
Figure~\ref{fig:G}a shows that the resistance of both the 2DEG and ohmic contacts exhibit a weak temperature dependence. 
These transport measurements were performed down to 1\,mK with no discernible heating using conventional room temperature electronics
without a screened room; this demonstrates the effectiveness of the filtering. A change in the resistance was observed between 0.4 and 0.8\,K, 
that has been identified in similar samples~\cite{Beauchamp2020} to be a superconducting transition in the ohmic contacts.

\textbf{Thermal transport measurements and thermal model.}
The use of an external thermometer in our setup requires an investigation of the heat flow in the system in order to relate
the electron temperature in the 2DEG $T_{\text e}$ to the 2DEG NT temperature $T_{\text{NT}}$.
We consider two sources of heat: $\dot Q_{\text{NT}} = \dot Q_{\text{NT}}^0 + \dot Q_{\text{NT}}^{\text J}$ applied to the NT
and $\dot Q_{\text e} = \dot Q_{\text e}^0 + \dot Q_{\text e}^{\text J}$ uniformly generated within the 2DEG device.
Each includes a residual heat leak ($\dot Q_{\text{NT}}^0$ and $\dot Q_{\text e}^0$);
additionally, to study the thermal response Joule heating $\dot Q_{\text{NT}}^{\text J}$ is applied using the NT heater
and $\dot Q_{\text e}^{\text J}$ by driving a current between I$+$ and I$-$ ohmic contacts.

\begin{figure}[t!]
\centerline{\includegraphics{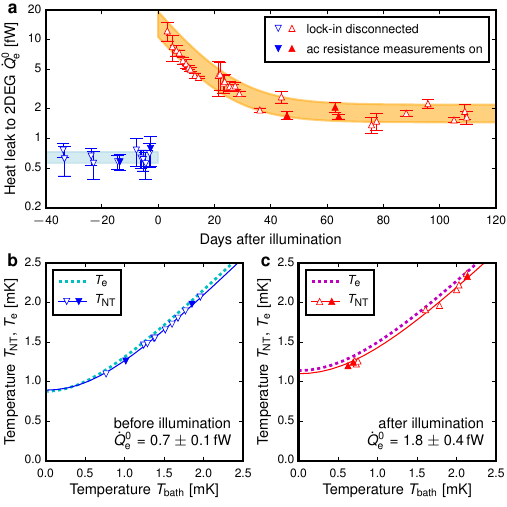}}
\caption{\textbf{Heat leak to the electrons in 2DEG and inferred electron temperature before (blue) and after (red) illumination.}
\textbf a~The heat leak $\dot Q_{\text e}^0$ to the 2DEG, determined from  Eq.~\eqref{eq:dotQ:Tnt}.
Error bars and thick lines through the data represent s.d.
Low-excitation AC transport measurements (filled triangles) do not increase the heat leak significantly
over the value it takes when no room-temperature electronics are connected to the experiment (open triangles).
After illumination with the LED the heat leak increases dramatically and then gradually decays, stabilising a month later;
the orange band that highlights the exponential decay is a guide to the eye.
\textbf{b},\textbf{c} The triangles represent the measurements of $T_{\text{NT}}$ vs $T_{\text{bath}}$ 
corresponding to the $\dot Q_{\text e}^0$ shown in \textbf{a}; the solid lines are fits to Eq.~\eqref{eq:dotQ:Tnt}.
The dotted lines show the electron temperature $T_{\text{e}}$ in the 2DEG estimated from Eq.~\eqref{eq:dotQ:Te}.
The first 40 days after illumination are excluded from \textbf{c} due to evolving $\dot Q_{\text e}^0$.}
\label{fig:dotQ:Te}
\end{figure}

Figure~\ref{fig:G}b shows the thermal conductance $G_\Sigma(T_{\text{NT}}) = d \dot Q_{\text{NT}} / d T_{\text{NT}}$
between the 2DEG NT and the \He3 bath, as inferred from measurements of $T_{\text{NT}}$ as a function of
$\dot Q_{\text{NT}}$ at a constant \He3 bath temperature $T_{\text{bath}}$.
Figure~\ref{fig:G}c illustrates the two parallel channels of $G_\Sigma = G_{\text e} + G_{\text{He}}$,
$G_{\text e}$ via electrons in the 2DEG and ohmic contacts, and the Kapitza boundary conductance $G_{\text{He}}$
between the NT assembly and the \He3 bath~\cite{Nakayama1989}.
The illumination of the 2DEG increases $G_{\text e}$, allowing us to separate it from $G_{\text{He}}$.
To reduce $G_{\text{He}}$ the heat exchangers in the immersion cell were plated
with approx.\ 30\,$\mu$mol/m$^2$ \He4 coverage before loading \He3~\cite{Hu1996}.

In the initial stage of this research we assumed that both the 2DEG and the ohmic contacts obey the Wiedemann-Franz (WF) law, 
ubiquitous in electronic transport in the $T\to 0$ limit~\cite{Sommerfeld1928,Syme1989,Kumar1993b,Chiatti2006,Jezouin2013,Macia2009}.
The WF law predicts that $G_{\text e}^{\text{WF}} = L_0 T/R_{\text{NT\,to\,IV}}$, where
$R_{\text{NT\,to\,IV}} = 27\,\Omega \  (6\,\Omega)$ is the electrical resistance from the NT ohmic contact
to I$\pm$ and V$\pm$ connected in parallel, measured before (after) illumination, where
$L_0 = \pi^2 k_{\text B}^2 / 3 e^2 = 2.44\times 10^{-8}\,\mathrm{W\Omega / K^2}$ is the Lorenz number~\cite{Sommerfeld1928}.
Figure~\ref{fig:G}b shows that $G_\Sigma < G_{\text e}^{\text{WF}}$ after illumination,  
a clear violation of the WF law that we attribute to superconductivity~\cite{Zavaritskii1958} in the ohmic contacts~\cite{Beauchamp2020}.
To the leading order the change in $G_\Sigma$ due to illumination is $\Delta G_\Sigma = \Delta G_{\text e} \propto T$;
to explain this observation we propose a model of partially-superconducting ohmic contacts, see Fig.~\ref{fig:G}d,
such that well below the superconducting transition temperature $T_c$,
$G_{\text e} = L_0 T / R_{\text e}$ is described by a resistance $R_{\text e} \ge R_{\text{NT\,to\,IV}}$.
The data are well described by
\begin{equation}
\label{eq:G}
G_\Sigma(T) = L_0 T \big / R_{\text e} + K_2 T^2 + K_3 T^3,
\end{equation}
with $R_{\text e} = 59\,\Omega$ ($27\,\Omega$) before (after) illumination and illumination-independent $G_{\text{He}} = K_2 T^2 + K_3 T^3$.
Similarly the response to $\dot Q_{\text e}$ was characterised in terms of the non-local thermal conductance
$G_{\text N}(T_{\text{NT}}) = d \dot Q_{\text e} / d T_{\text{NT}}$, see Fig.~\ref{fig:G}e, which was found to follow
\begin{equation}\label{eq:GN}
G_{\text N}(T) = \big(G_\Sigma(T) + X_4 T^4\big) \big/\alpha,
\end{equation}
with an illumination-independent term $X_4$,
but with different values of $\alpha$ before and after illumination.
The $X_4 T^4$ term describes an additional cooling mechanism in the 2DEG or ohmic contacts, 
insignificant below $T_{\text{NT}} = 3$\,mK, where $X_4 T^4 \ll G_\Sigma(T)$. 
In this low temperature regime, Eq.~\eqref{eq:GN} reduces to $G_{\text N} = G_\Sigma / \alpha$ 
where $\alpha$ is the fraction of $\dot Q_{\text e}$ that flows towards the NT ohmic contact, as shown in Figs.~\ref{fig:G}f,g.
Then
\begin{align}
\dot Q_{\text{NT}} &+ \alpha\dot Q_{\text e} = \int\limits_{T_{\text{bath}}}^{T_{\text{NT}}} \!\! G_\Sigma(T)\, dT
= \frac{L_0}{2 R_{\text e}}\big(T_{\text{NT}}^2 - T_{\text{bath}}^2\big)\quad\notag\\[-0.2em]
	&+ \frac{K_2}{3}\big(T_{\text{NT}}^3 - T_{\text{bath}}^3\big)
	+ \frac{K_3}{4}\big(T_{\text{NT}}^4 - T_{\text{bath}}^4\big).
\label{eq:dotQ:Tnt}
\end{align}
The combined heat leak $\dot Q_{\text{NT}}^0 + \alpha\dot Q_{\text e}^0$ can be inferred
from the difference between $T_{\text{NT}}$ and $T_{\text{bath}}$ in the absence of Joule heating.
We assume $T_{\text{bath}}$ to be equal to the fridge temperature, as justified in SI.
By replacing the gold thermal link between the NT ohmic contact and the NT with an aluminium one, 
we measured $\dot Q_{\text{NT}}^0 = 0.08 \pm 0.02$\,fW in a separate experiment (see SI),
allowing us to obtain $\dot Q_{\text e}$ directly, see Fig.~\ref{fig:dotQ:Te}a.
Before illumination by the LED the heat leak was found to be $\dot Q_{\text e}^0 = 0.7\pm 0.1$\,fW.
After illumination there was a dramatic increase of $\dot Q_{\text e}^0, $
followed by a slow relaxation, consistent with slow recombination processes in the heterostructure~\cite{Lin1990}.
On a time scale of a month, a new stable level $\dot Q_{\text e}^0 = 1.9\pm 0.4$\,fW was reached,
higher than before the illumination.

To estimate the minimum electron temperature in the 2DEG we use the simple model shown in Fig.~\ref{fig:G}f, 
in which $\dot Q_{\text e}$ is a point source of heat in the middle of the device
and additional cooling associated with the $X_4 T^4$ term in Eq.~\eqref{eq:GN} is ignored:
\begin{equation}\label{eq:dotQ:Te}
T_{\text e}^2 = (1-\alpha) \, T_{\text{bath}}^2 + \alpha \, T_{\text{NT}}^2
+ 2 \alpha (1-\alpha) \, R_{\text e} \dot Q_{\text e} / L_0.
\end{equation}
Figures~\ref{fig:dotQ:Te}b and c show the inferred electron temperature $T_{\text e}$, demonstrating that
we have cooled the electrons to 1\,mK, despite the unexpected hindrance posed by the superconductivity in the ohmic contacts.

\section{Discussion}
We demonstrate that at 1\,mK the dominant cooling of the 2D electrons is by the ohmic contacts,
which have a thermal conductance of $G_{\text e} \sim 10^{-9}\,$W/K$^2 \times T$.
The electron-phonon coupling in high-mobility 2DEGs has been observed~\cite{appl98a} to follow
$\dot Q_{\text{e-ph}} = 61\,$eV/s\,K$^5\times N(T_{\text e}^5 - T_{\text{ph}}^5)$
at phonon temperatures of $T_{\text{ph}} = 0.3$-0.5\,K, where $N$ is the number of electrons.
For our device this leads to a thermal conductance
$G_{\text{e-ph}} = d\dot Q_{\text{e-ph}} / dT_{\text e} \sim 10^{-6}$\,W/K$^5\times T^4$.
Assuming this high-temperature power law extrapolates to ULT, the cross-over $G_{\text e} = G_{\text{e-ph}}$
occurs near 100\,mK and at 1\,mK $G_{\text{e-ph}} / G_{\text e} = 10^{-6}$, 
so cooling via electron-phonon coupling alone would only achieve $T_{\text e} = 20\,$mK (70\,mK) for
$T_{\text{ph}} = 0.3$\,mK and $\dot Q_{\text e}^0 \sim 1\,$fW (500\,fW).

The $X_4 T^4$ term in Eq.~\eqref{eq:GN} points towards an additional cooling channel competing with $G_{\text e}$ above 3\,mK,
such as enhanced $G_{\text{e-ph}} \gg 10^{-6}$\,W/K$^5\times T^4$ at ULT;
alternatively $G_{\text{e}}(T)$, $G_{\text{NT}}(T)$ and $G_{\text{IV}}(T)$, see Fig.~\ref{fig:G}f,
may deviate from the $G \propto T$ behaviour, if the effective contact resistance $R_c^*$ (see Fig.~\ref{fig:G}d) is not constant at ULT.
Ignoring these effects in Eq.~\eqref{eq:dotQ:Te} potentially overestimates $T_{\text e}$ for the preliminary experiments
shown in Fig.~\ref{fig:immcells}c, but the ultimate electron temperature we report
and the qualitative observation of the reduction of the heat leak $\dot Q_{\text e}$ due to the electromagnetic shielding are robust.
In addition to the $X_4 T^4$ term, a more detailed thermal model, beyond the scope of this work, should take
into account the distribution of $\dot Q_{\text e}$ across the 2DEG and 2D and 3D heat flow in the 2DEG and ohmic contacts.

We note that after illumination the thermal resistances are dominated by the ohmic contacts,
since $R_{\text e} \gg R_{\text{NT\,to\,IV}}$.
Therefore the thermal conductance through the I$\pm$ and V$\pm$ contacts in parallel,
$G_{\text{IV}} = G_{\text e}/\alpha$, (see Fig.~\ref{fig:G}f), 
is only $(1-\alpha)/\alpha = 2.1$ times higher
than that through the NT ohmic, $G_{\text{NT}} = G_{\text e}/(1-\alpha)$, 
despite being 15 times larger in circumference.
This suggests that the large I$\pm$ ohmics were rendered thermally inactive
due to low thermal conductance along them,
as the gold wires were bonded 0.5\,mm away from their front edges, Fig.~\ref{fig:metalcell}e,
to prevent damage to the 2DEG adjacent to these ohmics.
The larger value of $(1-\alpha)/\alpha = 3.2$ before illumination is consistent
with the I$\pm$ ohmics being more active when the 2DEG resistance is higher.

A natural question to ask is whether even lower electron temperatures can be achieved.
First, the heat leak may be reduced by optimising the device geometry
(for some sources $\dot Q_{\text e}^0 \propto \text{area}$) and further filtering.
We recognise that our NT readout scheme limited the filters between
the NT and the SQUID to a two-way Ag epoxy filter with a high cut-off frequency.
Improved electron cooling is expected in experiments compatible with heavy filtering of every measurement line.

Another approach is to improve the ohmic contacts. Their thermal conductance can be increased if the superconductivity
is suppressed with a magnetic field or by a change of recipe.
To optimise the performance of the large partially-superconducting ohmic contacts used in this work, a thick gold film could be evaporated
on the annealed top surface or a dense network of gold wires could be bonded along the front edge of the contacts.
We estimate electron temperatures of 0.4-0.6\,mK, if the above steps are combined with the optimised fridge performance (see SI).

Important future steps include characterising and mitigating the heating associated with surface gates
used to define mesoscopic samples, measuring the electron temperature in the device directly~\cite{Iftikhar2016,Kleinbaum2017},
and extending the techniques presented here to high magnetic fields. 
Based on the thermal measurements at 6-100\,mK~\cite{Iftikhar2016}, 
the quantum Hall states will cool into the microkelvin regime in our environment.
The immersion cooling can efficiently thermalise a variety of degrees of freedom in condensed matter systems,
offering a promising path to control the decoherence problem in superconducting electronics~\cite{deGraaf2017}.

In conclusion, the demonstrated cooling of a large area two-dimensional electron system to below 1\,mK constitutes
a significant technical breakthrough in quantum nanoelectronics.
Fundamental studies of lower dimensional electron devices such as quantum wires and quantum dots in the microkelvin regime
can be achieved if the heat leak to the 2DEG, generated by the surface gates, is kept at the fW level.   
Likewise, the addition of a strong magnetic field, which is straightforward to implement in our modular design, 
coupled with samples of the highest quality, is poised to contribute to the understanding of novel ground states
of two-dimensional electron systems at fractional LL fillings, arising from CF interactions. 
Thus the microkelvin regime is opened up for the study of a rich and diverse array of strongly correlated quantum systems,
with high potential for future discovery.

\section{Methods}

\textbf{2DEG sample.}
The 4\,mm$\times$4\,mm sample shown in Fig.~\ref{fig:immcells}b was fabricated
using wafer W476; details of its MBE growth and the subsequent fabrication of AuNiGe ohmic contacts are given elsewhere~\cite{Beauchamp2020},
together with the critical field measurements of the superconducting state below 1~K.
The as-grown 2D carrier density and mobility are $n_{\text{2D}} \approx 2 \times 10^{11}$\,cm$^{-2}$ and $\mu \approx 1 \times 10^6$\,cm$^2$/Vs
giving a calculated sheet resistance of $R_\Box \approx 31\,\Omega$.
After illumination these quantities are $3.3 \times 10^{11}$\,cm$^{-2}$, $3 \times 10^6$\,cm$^2$/Vs
and $6\,\Omega$. 
Illumination was performed at 1.5\,K with the immersion cell evacuated.
The LED was driven with currents up to 2\,mA for 1 minute, 
by which time $R_\Box$ saturated.

\textbf{Filters.}
The Ag epoxy filters, shown in Fig.~\ref{fig:metalcell}d, combine features of several designs~\cite{Scheller2014,Bluhm2008,Tancredi2014}.
Metre-long superconducting NbTi wires were coiled around threaded sterling silver tubes and encapsulated in conducting Ag epoxy.
In the resulting lossy coaxial the dissipation occurs in the outer conductor which is in direct metallic contact with the refrigerator,
ensuring good thermalisation. These filters with a 100\,MHz cut-off were combined with the Cu powder filters with 2\,MHz cut-off, 
and the 16\,kHz low-pass $RC$ filters shown in Fig.~\ref{fig:metalcell}b.
To prevent high-frequency leaks the Ag epoxy filters were connected to the immersion cell via semi-rigid coaxial cables with threaded connectors.

A version of the Ag epoxy filter using a NbTi twisted pair (instead of a single wire)
embedded in Ag epoxy was inserted between 2DEG NT and its SQUID sensor. This design was chosen for having zero DC resistance and
for inducing sufficiently low amounts of noise in the SQUID so as to not mask the NT signal.
This filter was mounted directly on top of the immersion cell and sealed
to the cell wall with In to maintain a photon-tight enclosure.
Another multi-line Ag epoxy filter at 1.5\,K was inserted between the SQUID sensor and its room temperature electronics.
All remaining filters were mounted at the mixing chamber plate of the dilution refigerator.

\textbf{Noise thermometry.}
Both NTs were read out with integrated 2-stage SQUID current sensors~\cite{Drung2007} 
and were calibrated against a primary magnetic field fluctuation thermometer at 10-100\,mK~\cite{Kirste2016}.

\textbf{Electrical transport measurements} were performed using a mains-powered Stanford Research Systems SR-124 analogue lock-in amplifier
using its internal oscillator. Below 20\,mK measurement currents between 1 and 10\,nA were used to ensure sub-fW Joule heating.

\begin{figure}[t!]
\includegraphics{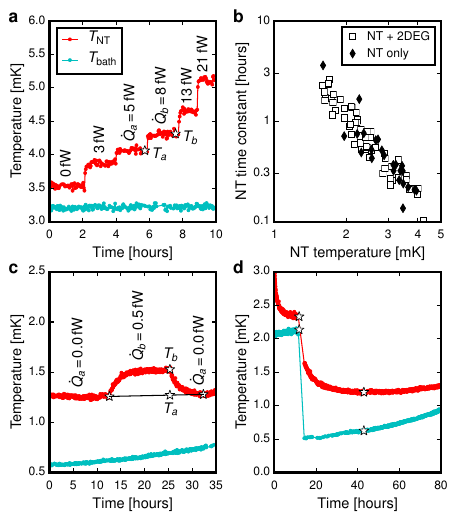}
\vskip-0.5em
\caption{\textbf{Measurements of the thermal conductance and heat leak.}
\textbf{a} The thermal conductance $G_{\Sigma}$ was inferred from the step-wise measurement
of $T_{\text{NT}}(\dot Q_{\text{NT}}^{\text J})$ when $T_{\text{bath}}$ was kept constant.
\textbf{b} The slow response of $T_{\text{NT}}$ to heaters and $T_{\text{bath}}$ below 3\,mK
required a modified $G_{\Sigma}$ measurement shown in \textbf{c}.
\textbf{d} Heat leak was inferred via stationary Eq.~\eqref{eq:dotQ:Tnt}
which applies when $dT_{\text{NT}}/dt = 0$.}
\vskip-1em
\label{fig:thermal}
\end{figure}

\textbf{Thermal measurements.}
The thermal conductance $G_\Sigma(T_{\text{NT}}) = d\dot Q_{\text{NT}} / d T_{\text{NT}}$ 
was inferred from pairs of measurements of $T_{\text{NT}} = T_{a,b}$ at different levels of 
$\dot Q_{\text{NT}}^{\text J} = \dot Q_{a,b}$
at constant $T_{\text{bath}}$ as 
$G_\Sigma\big((T_a + T_b)/2\big) = \big(\dot Q_b - \dot Q_a\big)\big/\big(T_b - T_a\big)$, Fig.~\ref{fig:thermal}a.
Identical techniques were used for measuring $G_{\text{N}}(T)$.
Below 3\,mK the 2DEG NT responded slowly, Fig.~\ref{fig:thermal}b, and $T_{\text{bath}}$ could not be kept constant
sufficiently long due to the single-shot operation of the demagnetisation cooling,
so $T_a$ was inferred from measurements before and after $T_b$, see Fig.~\ref{fig:thermal}c.
Since similar slow relaxation was observed when NT was isolated from the 2DEG, see Fig.~\ref{fig:thermal}b,
we conclude that its origin is the large heat capacity of the NT itself, and all other elements of the system thermalise faster.
This implies that when $dT_{\text{NT}}/dt = 0$, see Fig.~\ref{fig:thermal}d,
the entire system is in a steady state, allowing the stationary Eq.~\eqref{eq:dotQ:Tnt} to be used to extract the heat leak.
The slow NT response practically limits the thermal measurements to $T_{\text{NT}} > 1.2$\,mK.

In the early experiments shown in Fig.~\ref{fig:immcells}c, $T_{\text{NT}}$ was too high 
to ignore the $X_4 T^4$ term in $G_{\text N}$. 
Here the thermal model given by Eqs.~\eqref{eq:dotQ:Tnt} and \eqref{eq:dotQ:Te} is inaccurate. 
The heat leak was inferred from
$\dot Q_{\text e}^0 = \int\limits_{T_{\text{bath}}}^{T_{\text{NT}}} G_{\text N}(T)\,dT$.
To account for the use of different 2DEG devices of the same design
and the lack of \He4 plating $G_{\text N}(T)$ was measured separately in each experiment, see SI for further information. 

\textbf{Data availability.} The thermal conductance, electrical resistance, heat leak and thermal time constant
data obtained in this work are available in Figshare at \mbox{\url{https://doi.org/10.6084/m9.figshare.17057063}}.


\begin{acknowledgments}
\textbf{Acknowledgements.} We thank P.~Bamford, R.~Elsom, 
I.~Higgs and J.~Taylor for mechanical support, and V.~Antonov for wire bonding.
Cu powder filters designed and fabricated by A.~Iagallo and M.~Venti were instrumental in the early stages of this work.
We acknowledge fruitful discussions with C.~Ford, R.~Haley, P.~Meeson, P.~See and J.~Waldie.
This research was supported by EPSRC Programme grant EP/K004077/1 and
the EU H2020 European Microkelvin Platform EMP, Grant No. 824109.
\end{acknowledgments}

{\small
\textbf{Author contributions.} 
JS, ADC, JN, AJC and LVL conceived the experiments;
the wafer was grown by DAR and IF; ADC, GC and JTN designed and fabricated the 2DEG device; 
ADC and LVL designed and constructed the plastic immersion cell, with input from JN, AJC, JTN and JS.
LVL and HvdV designed and constructed the metal immersion cell and filters
and performed ultra-low temperature measurements in both cells, with input from JN, AJC, JTN and JS;
SD and LVL measured the heat leak to the 2DEG;
TT, ML, AJC and JTN characterised the superconductivity in the ohmic contacts;
LVL analysed the data; LVL, JS and JTN wrote the manuscript with critical input from all authors.

\textbf{Competing interests.} The authors declare no competing interests.}

\onecolumngrid
\pagebreak

\renewcommand{\thefigure}{S\arabic{figure}}
\renewcommand{\thetable}{S\arabic{table}}
\renewcommand{\theequation}{S\arabic{equation}}
\setcounter{figure}{0}
\setcounter{table}{0}
\setcounter{equation}{0}

\begin{center}
{\large\textbf{Cooling Low-Dimensional Electron Systems into the Microkelvin Regime\\
Supplementary Information}}\\[1em]
Lev V. Levitin,$^1$ Harriet van der Vliet,$^1$ Terje Theisen,$^1$ Stefanos Dimitriadis,$^1$\\
Marijn Lucas,$^1$ Antonio D. Corcoles,$^1$ J\'{a}n Ny\'{e}ki,$^1$ Andrew J. Casey,$^1$ Graham Creeth,$^2$\\
Ian Farrer,$^3$ David A. Ritchie,$^3$ James T. Nicholls,$^1$ and John Saunders$^1$\\[0.5em]
{\small\textit{$^1$ Department of Physics, Royal Holloway, University of London, Egham TW20 0EX, UK}\\
\textit{$^2$ London Centre for Nanotechnology, University College London, London WC1H 0AH, UK}\\
\textit{$^3$ Cavendish Laboratory, University of Cambridge, JJ Thomson Avenue, Cambridge CB3 0HE, UK}\\
\makeatletter
(\@date)}
\makeatother
\end{center}

\vskip1em

\twocolumngrid

\begin{figure}[b!]
\centerline{\includegraphics{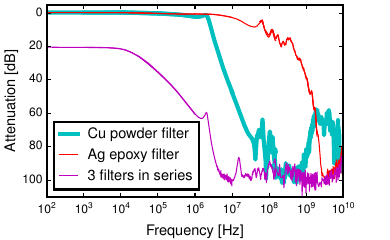}}
\caption{{Attenuation characteristics of the different low-pass filters at 4.2\,K.} The Cu powder filter,
shown in Fig.~\ref{fig:metalcell}c, transmits near 100\,MHz and 3\,GHz, above its 2\,MHz cut-off frequency.
The Ag epoxy filter with a single superconducting core, see Fig.~\ref{fig:metalcell}d,
has a 100\,MHz cut-off, higher than in Ref.~\cite{Scheller2014} due to the lack of dissipation in the superconducting core.
The multi-line Ag epoxy filter used between the 2DEG NT and its SQUID sensor exhibits similar attenuation for
the common mode signals. The differential mode, which has not been characterised, may show less attenuation.
A chain of 3 filters: single-line Cu powder and Ag epoxy filters plus an $R=500\,\Omega$, $C=20$\,nF filter
(a pair of Fig.~\ref{fig:metalcell}b filters wired in parallel, such pairs were used in the NT heater lines to
reduce mixing chamber heating when driving large currents through the heater),
has a 16\,kHz cut-off determined by the $RC$ filter and at least $100$\,dB attenuation at 0.03-3\,GHz.
The ``lossy coaxial'' design of the Ag epoxy filters~\cite{Tancredi2014} and the heterogeneity
of the chain components makes the transmission above the cut-off frequency, as exhibited by the Cu powder filter, unlikely.
None of the filters match the 50\,$\Omega$ impedance of the spectrum analyser and vector network analyser
used in this characterisation. This mismatch potentially results in an overestimated attenuation.}\label{fig:filters}
\vskip-0.45em
\end{figure}

\section{Further details of the~experimental~setup}
All experiments were conducted on a home-made copper nuclear demagnetisation stage,
precooled by a custom-built ``wet'' Oxford Instruments Kelvinox 400 dilution refrigerator.
The metal cell was constructed out of OFHC copper and high purity silver, with small amounts of niobium,
copper-nickel, silicon and amorphous dielectrics, such as PEEK, Teflon, Stycast 1266, paper and GE varnish.
The heat exchangers were fabricated from 70\,nm silver powder.
Following Ref.~\cite{Scheller2014} silver epoxy filters, Fig.~\ref{fig:metalcell}d, were made using Epotek E4110 conducting epoxy.
This epoxy was also used to attach the 4$\times$4\,mm 2DEG device to a copper holder that was screwed to the immersion cell lid,
to ensure good thermalisation of phonons in GaAs.
The plastic cell was made out of Stycast 1266 epoxy and included the materials listed above.

Figure~\ref{fig:filters} shows the 4.2\,K attenuation characteristics of the low-pass filters used in this work.

\section{NT heater resistance}
\begin{figure}[h!]
\centerline{\includegraphics{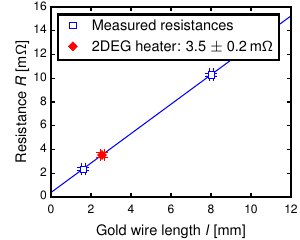}}
\caption{The resistance of the 2DEG NT heater is inferred from its length from $R(l)$ measurements
on similar wires of different lengths, all connected with wedge-bonded superconducting aluminium wires.
The non-zero intercept in the $R = a+bl$ fit is attributed to the gold-aluminium contact resistance.
Error bars represent s.d.}
\label{fig:Rheater}
\end{figure}

The NT heater resistance, which determines the magnitude of $\dot Q_{\text{NT}}^{\text J}$, could not be measured in situ,
and was inferred from $R(l)$ measurements of similar wires of different lengths $l$ shown in Fig.~\ref{fig:Rheater}.

\section{2DEG resistance measurements}

\begin{figure}[t!]
\centerline{\includegraphics[width=3.1in]{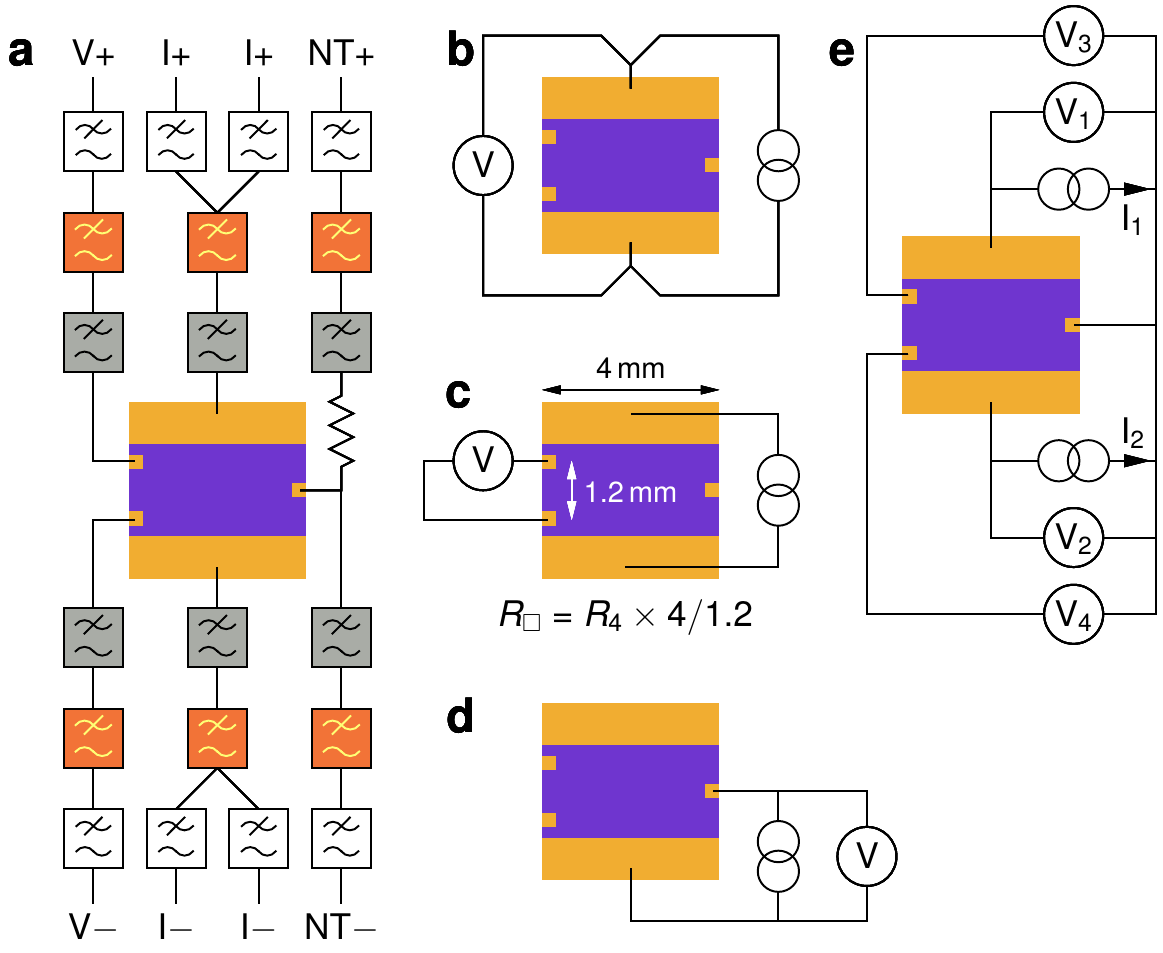}}
\caption{Details of the electrical measurements. \textbf{a}~Electrical connections to the 4\,mm$\times$4\,mm 2DEG device in the metal cell,
allowing 4-terminal resistance measurements using the NT and I$\pm$ ohmic contacts.
\textbf{b-d}~Measurements of $R_{\text{I$+$\,to\,I$-$}}$ (\textbf{b}), $R_\Box$ (\textbf{c})
and $R_{\text{NT\,to\,I$-$}}$ (\textbf{d}).
The 2DEG sheet resistance $R_\Box$ was obtained from the measured 4-terminal resistance $R_4$ by simple geometrical scaling.
\textbf{e} Measurement of $R_{\text{NT\,to\,IV}}$. 
Currents $I_1$ and $I_2$ fed into the I$\pm$ ohmics were adjusted to equalise their 
potentials $V_1$ and $V_2$, thus giving the resistance from the NT
to the I$\pm$ ohmic contacts to be  $R_{\text{NT\,to\,I$\pm$}} = V_1 / (I_1 + I_2)$.
Simultaneously $V_3 \approx V_4 \approx 0.99 V_1$, demonstrating
that $R_{\text{NT\,to\,IV}} \approx R_{\text{NT\,to\,I$\pm$}}$.}
\label{fig:electrical:SI}
\end{figure}

\begin{figure}[t!]
\centerline{
  \includegraphics{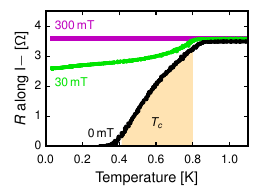}~~~~
  \raisebox{1em}{\includegraphics[width=0.5in]{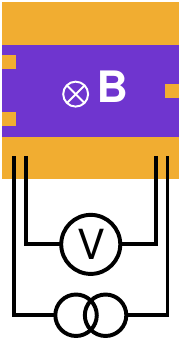}}}
\caption{Superconducting behaviour in the I$-$ ohmic contact (left), 
showing the suppression of $T_c$ when a 
magnetic field $\vec B$ is applied perpendicular to the 2DEG. 
These resistance measurements along the top of the contact were performed on the 4\,mm$\times$4\,mm 
device on a separate cool-down using a different wire bond configuration (right).}
\label{fig:RIminus}
\end{figure}

In addition to the filtering shown in Fig.~\ref{fig:metalcell}, extra $RC$ filters were connected to I$+$ and I$-$ lines
in the metal cell to provide independent current and voltage lines, as shown in Fig.~\ref{fig:electrical:SI}a.
Therefore the four-terminal resistance measurements that involve I$\pm$ contacts, such as $R_{\text{NT\,to\,I$-$}}$,
include Ag epoxy and Cu powder filters and the interconnecting Cu and NbTi wires, contributing less than 0.1\,$\Omega$.
Figure~\ref{fig:electrical:SI}e shows that a measurement of $R_{\text{NT\,to\,IV}}$
does not require shorting the I$\pm$ and V$\pm$ ohmics. Signatures of superconductivity in the I$-$ ohmic contact are shown in Fig.~\ref{fig:RIminus}.
Table~\ref{tab:R} summarises all the relevant low temperature resistances.

A different sample of identical design, also fabricated using wafer W476, was investigated in the plastic cell.
The LED was situated outside the cell and the light had to travel through the cell wall,
leading to less efficient illumination, see Table~\ref{tab:R:plastic},
than in the metal cell. Single I$\pm$ connections increased the uncertainty in $R_{\text{I$+$\,to\,I$-$}}$.
The absence of NT$\pm$ lines made in-situ measurements involving the NT ohmic contact impossible.

\begin{table}[t!]
\caption{{Low temperature resistances in the 4\,mm$\times$4\,mm device used in the metal cell.}
The resistance $R_{\text{I$_+$\,to\,I$_-$}}$ determines the magnitude of the 
Joule heating $\dot Q_{\text e}^{\text J}$.
The resistance from the NT ohmic contact to all other contacts connected in parallel, $R_{\text{NT\,to\,IV}}$,
determines the thermal conductance $G_{\text e}^{\text{WF}}$ predicted from the WF law.}
\label{tab:R}
\vskip0.5em
\centerline{\begin{tabular}{@{\hskip2em}c@{\hskip2em}c@{\hskip2em}c@{\hskip2em}}
  \hline
  \hline
  \rule{0pt}{1.1em} & \textbf{Before} & \textbf{After}\\
                    & \textbf{illumination} & \textbf{illumination}\\
  \hline
  \rule{0pt}{1.1em}
  $R_\Box$                     & 32.5\,$\Omega$ & 6.4\,$\Omega$ \\
  $R_{\text{I$+$\,to\,I$-$}}$  & 16.8\,$\Omega$ & 4.4\,$\Omega$ \\
  $R_{\text{NT\,to\,I$+$}}$    & 31.0\,$\Omega$ & 7.7\,$\Omega$ \\
  $R_{\text{NT\,to\,I$-$}}$    & 31.4\,$\Omega$ & 7.7\,$\Omega$ \\
  $R_{\text{NT\,to\,IV}}$      & 26.8\,$\Omega$ & 6.1\,$\Omega$ \\
  \hline
  \hline
\end{tabular}}
\vskip1em
\caption{{Low temperature resistances in the 2DEG device used in the plastic cell.}}\label{tab:R:plastic}
\vskip0.5em
\centerline{\begin{tabular}{@{\hskip2em}c@{\hskip2em}c@{\hskip2em}c@{\hskip2em}}
  \hline
  \hline
  \rule{0pt}{1.1em} & \textbf{Before} & \textbf{After}\\
                    & \textbf{illumination} & \textbf{illumination}\\
  \hline
  \rule{0pt}{1.1em}
  $R_\Box$                     & 48.0\,$\Omega$ & 15.3\,$\Omega$ \\
  $R_{\text{I$+$\,to\,I$-$}}$  & $22\pm 1\,\Omega$ & $7\pm 1\,\Omega$ \\
  \hline
  \hline
\end{tabular}}
\end{table}

\begin{figure}[b!]
\centerline{\raisebox{0.05in}{\includegraphics[height=1.55in]{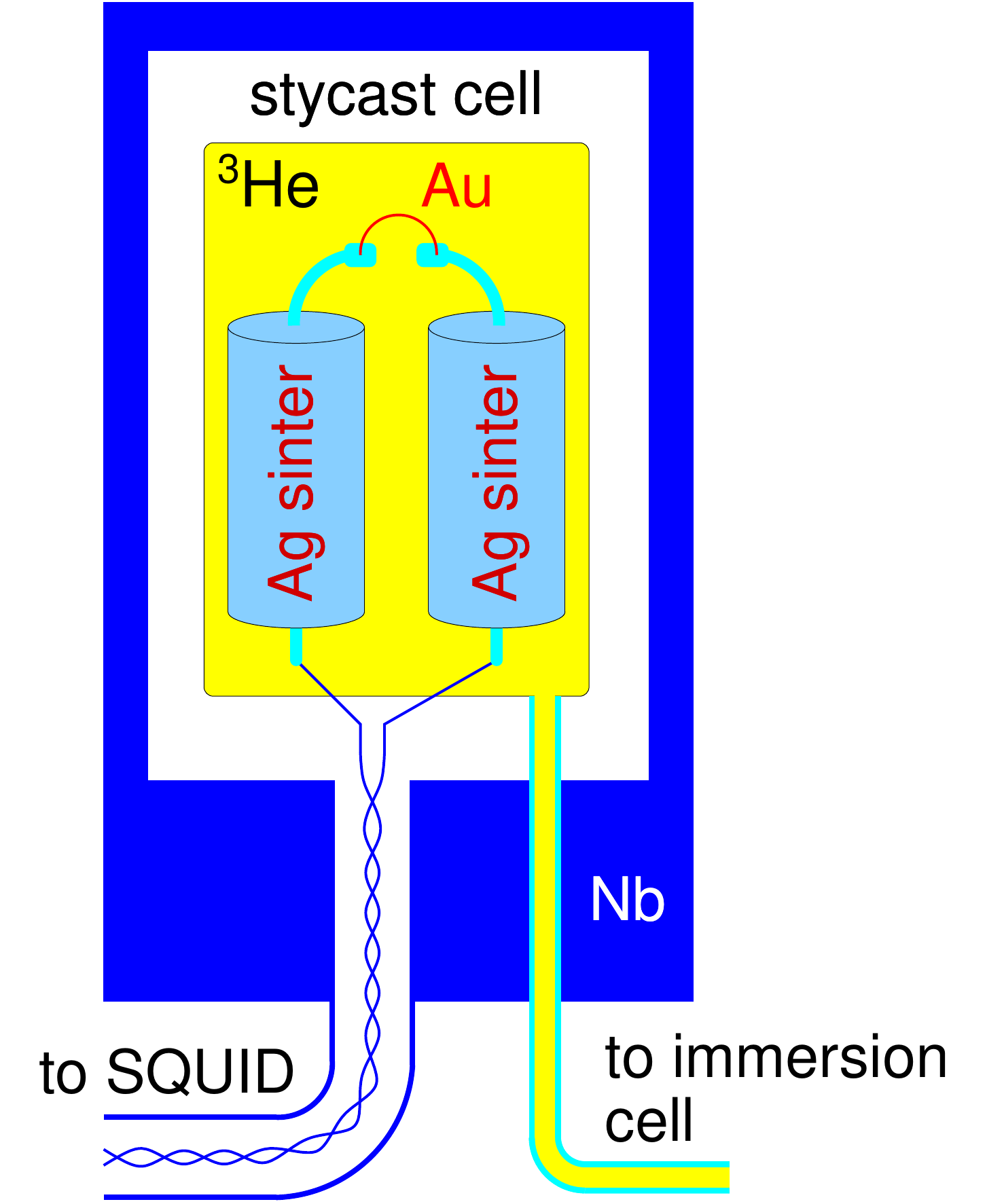}}\!\!\includegraphics{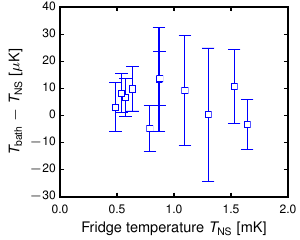}}
\caption{Thermalisation of \He3. In the plastic cell $T_{\text{bath}}$ was measured directly using a \He3 noise thermometer appendix (left).
It agrees with the noise thermometer monitoring the nuclear stage temperature $T_{\text{NS}}$ to within 20\,$\mu$K (right).
Error bars represent s.e.m.}\label{fig:T3vsTbath}
\vskip-0.45em
\end{figure}

\section{\He3 temperature}

The plastic immersion cell was equipped with a \He3 noise thermometer, see Fig.~\ref{fig:T3vsTbath}.
A 25\,$\mu$m Au wire was wedge-bonded between two sintered 1\,mm Ag wires. The whole assembly
was enclosed in a leak-tight Stycast 1266 appendix cell, connected to the main cell with a capillary and surrounded with a Nb shield.
The temperature $T_{\text{bath}}$ of this composite Au+Ag resistor was read out with a SQUID current sensor.
It was found to be within 20\,$\mu$K of the nuclear stage temperature $T_{\text{NS}}$ down to 0.5\,mK, see Fig.~\ref{fig:T3vsTbath}.

A similar thermometer was not implemented in the metal cell, to avoid the construction overheads
associated with bringing an additional twisted pair into the shielded environment. The low-heat leak nature of this environment
plus significantly reduced amount of amorphous dielectrics in contact with \He3, in comparison with the plastic cell,
allow us to conclude that the thermalisation of the \He3 bath in this metal cell was at least as good as in the plastic cell.
Therefore the temperature of the nuclear stage, monitored in the experiment with the metal cell,
gives an accurate measure of the \He3 bath temperature $T_{\text{bath}}$.

\section{2DEG Noise thermometer correction}
\begin{figure}[h!]
\centerline{\includegraphics{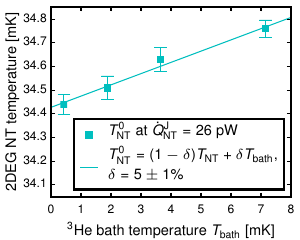}}
\caption{{The fraction $\delta$ of the 2DEG NT resistance sitting at the \He3 bath temperature $T_{\text{bath}}$.}
Here we present the raw 2DEG NT readings $T_{\text{NT}}^0$ versus $T_{\text{bath}}$ when high power is applied to the NT heater,
such that the temperature $T_{\text{NT}}$ of the NT gold wire is independent of $T_{\text{bath}}$.
A fit based on Eq.~\eqref{eq:NT0} allows us to determine $\delta$. Error bars represent s.d.}\label{fig:NTcoldfraction}
\end{figure}

The sintered silver heat sinks at $T_{\text{bath}}$, see Fig.~\ref{fig:metalcell}, 
installed to reduce $\dot Q_{\text{NT}}^0$,
contribute $\delta = 5$\% of the 2DEG NT resistance $R$, see Fig.~\ref{fig:NTcoldfraction}.
The measured Johnson voltage noise power
\begin{equation}\label{eq:NT0}
S_V = 4 k_B (1 - \delta) R T_{\text{NT}} + 4 k_B \delta R T_{\text{bath}} = 4 k_B R T_{\text{NT}}^0
\end{equation}
gives the raw NT reading $T_{\text{NT}}^0$, allowing us to infer the temperature $T_{\text{NT}}$ of the NT gold wire
\begin{equation}\label{eq:NTcorr}
T_{\text{NT}} = \frac{T_{\text{NT}}^0 - \delta T_{\text{bath}}}{1 - \delta}.
\end{equation}
The sub-$20\,\mu$K gradient between the \He3 bath temperature
and the measured nuclear stage temperature down to 0.3\,mK results in a negligible 1\,$\mu$K error in $T_{\text{NT}}$.

\begin{figure*}[t!]
\centerline{\includegraphics{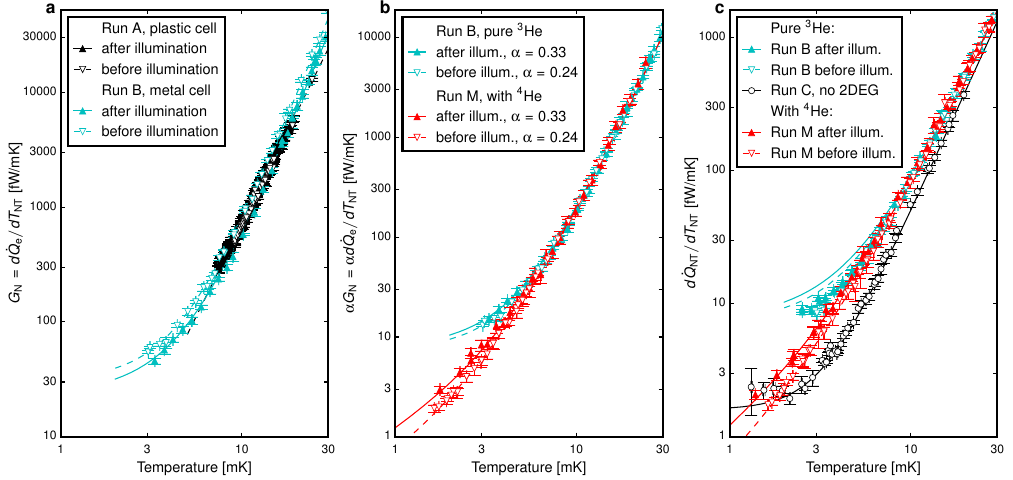}}
\centerline{\includegraphics{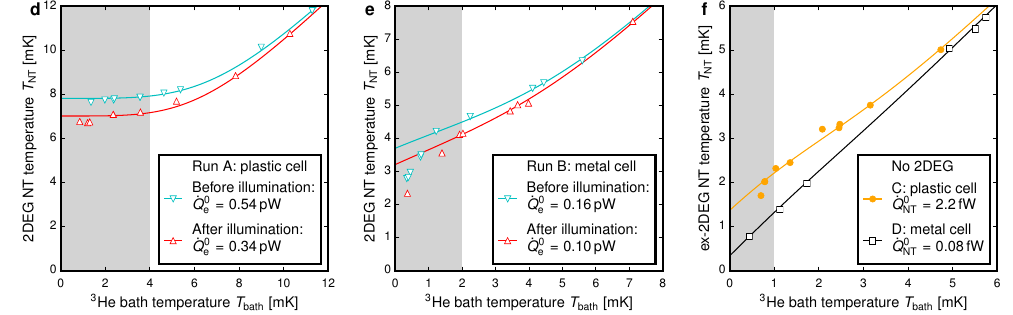}}
\caption{Thermal conductances and heat leaks in preliminary and auxiliary experiments.
\textbf{a} Non-local thermal conductance $G_{\text N}$ measured on two different 2DEG devices
in the preliminary runs A and B, both with pure \He3.
The data are described by Eqs.~\eqref{eq:GN:A} and \eqref{eq:GN:B}.
\textbf{b} $G_{\text N}$ measured on the same 2DEG device in the metal cell during runs B and M,
described by Eqs.~\eqref{eq:GN:B} and \eqref{eq:GN}. The data are shown scaled by $\alpha$ measured in run M.
\textbf{c}~Response of 2DEG NT to NT heater in runs B, C and M.
In run~M it provides a measurement of $G_\Sigma(T_{\text{NT}})$, Eq.~\eqref{eq:G},
however in run B the measured $d \dot Q_{\text{NT}} / T_{\text{NT}}$ (cyan triangles)
deviate from the predicted behaviour, see Eq.~\eqref{eq:G:B} (cyan lines) below 7\,mK,
attributed to significant $\dot Q_{\text e}^0$.
The Kapitza conductance between the NT assembly and \He3 measured in run~C 
is described by Eq.~\eqref{eq:GHe:C}.
\textbf{d-e}~Heat leak to 2DEG measured in runs A and B evaluated from $G_{\text{N}}$ in \textbf{a}.
\textbf{f}~Heat leak to NT + gold thermal link + NT heater in run~C and to NT alone in run~D.
The grey areas in \textbf{d}-\textbf{f} indicate the low temperature regime into which
$G_{\text{N}}(T)$ and $G_{\text{He}}(T)$ can not be reliably extrapolated.
These were excluded from the heat leak evaluation.
Error bars in \textbf{a}-\textbf{c} represent s.d.}\label{fig:SI:thermal}
\end{figure*}

\section{Additional thermal transport and heat leak measurements}
Figure~\ref{fig:SI:thermal} illustrates measurements of thermal conductances and heat leaks in the preliminary and
auxiliary immersion cell runs. In contrast to the main experimental run, referred to below as M,
in all of these measurements the immersion cells were filled with pure \He3,
resulting in 2D solid \He3 adsorbed on all surfaces inside the cells and consequently increased $G_{\text{He}}$~\cite{Hu1996}.
For each NT and NT heater the NT correction factor $\delta$ and heater resistance were obtained using procedures described above.
The coefficients in thermal conductance models obtained in the fits are summarised in Table~\ref{tab:thermal}.

\begin{table}[t!]
\vskip-0.55em
\caption{Summary of the fit parameters in Eqs.~\eqref{eq:G}, \eqref{eq:GN}, \eqref{eq:GN:A}, \eqref{eq:GN:B}, and \eqref{eq:GHe:C}.}
\label{tab:thermal}
\vskip0.5em
\centerline{\begin{tabular}{@{\hskip2em}c@{\hskip2em}c@{\hskip2em}c@{\hskip2em}}
  \hline
  \hline
  \rule{0pt}{1.1em} & \textbf{Before} & \textbf{After}\\
                    & \textbf{illumination} & \textbf{illumination}\\
  \hline
  \rule{0pt}{1.1em}%
$R$       & $59\,\Omega$       & $27\,\Omega$\\
$\alpha$  & 0.24               & 0.33\\
$K_0$     & \multicolumn{2}{c}{$7.0\times 10^{-12}\,$W/K$^{\phantom{3.2}}$}\\
$K'_0$    & \multicolumn{2}{c}{$1.6\times 10^{-12}\,$W/K$^{\phantom{3.2}}$}\\
$K_2$     & \multicolumn{2}{c}{$2.7\times 10^{-7\phantom{1}}\,$W/K$^{3\phantom{.2}}$}\\
$K_3$     & \multicolumn{2}{c}{$4.9\times 10^{-5\phantom{1}}\,$W/K$^{4\phantom{.2}}$}\\
$K_{3.2}$ & \multicolumn{2}{c}{$1.8\times 10^{-3\phantom{1}}\,$W/K$^{4.2}$}\\
$X_4$     & \multicolumn{2}{c}{$1.1\times 10^{-2\phantom{1}}\,$W/K$^{5\phantom{.2}}$}\\
  \hline
  \hline
\end{tabular}}
\end{table}

\textbf{Run A.} A 4\,mm$\times$4\,mm 2DEG device was cooled in the plastic immersion cell with Cu powder filters.
The sample was damaged during disassembly after the run, and so a second sample from wafer W476 and with similar electrical characteristics
was used in the metal cell. 
Figure~\ref{fig:SI:thermal}a shows the non-local thermal conductance measured between 7 and 30\,mK, which was found to follow
\begin{equation}\label{eq:GN:A}
G^{\text A}_{\text N}(T) = K_{3.2} T^{3.2},
\end{equation}
behaviour that is consistent with a sum of $T^3$ and $T^4$ terms
(see Eqs.~\eqref{eq:GN} and \eqref{eq:GN:B})
observed in runs M and B.
Unlike those runs, $G^{\text A}_{\text N}(T)$ is independent of the illumination, 
which may reflect a more favourable layout of the gold wire bonds to the I$\pm$ ohmics in run A.
The heat leak was inferred from measurements of $T_{\text{NT}}$ vs $T_{\text{bath}}$ at $\dot Q_{\text e}^{\text J} = 0$,
Fig.~\ref{fig:SI:thermal}d, and integrating Eq.~\eqref{eq:GN:A}
\begin{equation}\label{eq:Qdot0:A}
\dot Q_{\text e}^0 = \int\limits_{T_{\text{bath}}}^{T_{\text{NT}}} G^{\text A}_{\text N}(T)\,dT.
\end{equation}
Equation~\eqref{eq:Qdot0:A} gave consistent results above $T_{\text{bath}} = 4\,$mK.
This was not the case at lower temperatures suggesting that $G^{\text A}_{\text N}(T)$
does not extrapolate to low temperatures according to Eq.~\eqref{eq:GN:A}.
Therefore the data at $T_{\text{bath}} < 4\,$mK, shown in grey in Fig.~\ref{fig:SI:thermal}d,
were excluded from the determination of $\dot Q_{\text e}^0$.
Due to the lack of NT heater $G_{\Sigma}$ and $\alpha$ could not be measured directly, so
$T_{\text e}$ in Fig.~\ref{fig:immcells}c was roughly estimated using the parameters
derived for the second 2DEG device, with the heat leak $\dot Q_{\text e}^0$ being the only input from Run A.

\textbf{Run B.} The second 2DEG device was cooled in the metal cell with Cu powder filters.
In this run $G_{\text N}$ was found to follow
\begin{equation}\label{eq:GN:B}
G^{\text B}_{\text N}(T) = G^{\text M}_{\text N}(T) + K_0/\alpha,
\end{equation}
see Fig.~\ref{fig:SI:thermal}a, with $G^{\text M}_{\text N}(T)$ and $\alpha$ obtained in run M, Eq.~\eqref{eq:GN}.
This corresponds to the thermal conductance between the NT and the \He3 bath
\begin{equation}\label{eq:G:B}
G^{\text B}_\Sigma(T) = G^{\text M}_\Sigma(T) + K_0,
\end{equation}
with $G^{\text M}_\Sigma(T)$ given by Eq.~\eqref{eq:G}. Here $K_0$ represents an extra channel
in $G_{\text{He}}$ due to the nuclear magnetism of 2D solid \He3 absorbed on the noise thermometer assembly
when the cell is filled with pure \He3 in run~B, but suppressed by \He4 plating in run~M~\cite{Hu1996}.
The discrepancy between $d \dot Q_{\text{NT}} / d T_{\text{NT}}$ measured in run B
and Eq.~\eqref{eq:G:B} below 7\,mK, see Fig.~\ref{fig:SI:thermal}c, is attributed to the breakdown of the
differential thermal conductance measurement $d \dot Q_{\text{NT}} / d T_{\text{NT}}$
in the presence of significant $\dot Q_{\text e}^0$, in the regime where $G_\Sigma / G_{\text N}$ is temperature-dependent.
The heat leak, Fig.~\ref{fig:SI:thermal}e, was measured by the same procedure as 
in run~A and the extrapolation of the measured
$G^{\text B}_{\text N}(T)$ was found to hold down to 2\,mK.

\textbf{Run C.} To measure $G_{\text{He}}$ separately from $G_{\text e}$,
an NT + gold thermal link + NT heater assembly was installed in the plastic cell,
similar to that used in the metal cell, but with no 2DEG and with the NT ohmic contact substituted
with a gold bonding pad on a silicon chip.
The total length of the NT + thermal link + heater gold wire chain was similar to the metal cell,
with a surface area of 0.8\,mm$^2$.
Figure~\ref{fig:SI:thermal}c shows the Kapitza conductance between this gold wire and \He3 that follows
\begin{equation}\label{eq:GHe:C}
G^{\text C}_{\text{He}}(T) = K'_0 + K_3 T^3,
\end{equation}
with the same $K_3$ as in Eq.~\eqref{eq:G}, but without the $K_2T^2$ term and with $K'_0$ smaller than $K_0$ observed
run B, Eq.~\eqref{eq:GN:B}.
We attribute this discrepancy to the strong thermal link between the 0.04\,mm$^2$ surface of the NT ohmic contact and \He3,
potentially related to the coupling between \He3 nuclear spins and magnetic Ni in the ohmic contact.
The heat leak, Fig.~\ref{fig:SI:thermal}f was obtained as
\begin{equation}\label{eq:Qdot:C}
\dot Q_{\text{NT}}^0 = \int\limits_{T_{\text{bath}}}^{T_{\text{NT}}} G^{\text C}_{\text{He}}(T) \, dT.
\end{equation}
Run~C provides the upper bound on $\dot Q_{\text{NT}}^0$ for runs~A and B
and justifies ignoring the small $\dot Q_{\text{NT}}^0$ 
when evaluating much larger $\dot Q_{\text e}^0$.

\textbf{Run D.} After the end of run M, the gold thermal link between NT ohmic and NT was replaced with aluminium.
Isolating the NT from the 2DEG and NT heater, allowed us to measure just the heat leak 
into the 1.6\,mm long NT, as opposed to run C, where the combined heat leak into 10\,mm long chain of gold wires was measured.
To obtain the heat leak, Fig.~\ref{fig:SI:thermal}f, we scale $G^{\text C}_{\text{He}}$
in Eq.~\eqref{eq:Qdot:C} by the wire length. 
This provides the lower bound on $\dot Q_{\text{NT}}^0$ for run M, 
and allows us to obtain an upper bound on $\dot Q_{\text e}^0$ and $T_{\text e}$, 
which is the main purpose of our thermal model.

\section{Potential improvements in cooling}

Here we consider the cooling of the 2DEG device characterised in the metal cell
in the absence of the superconductivity in the ohmics, while retaining all other parameters.
We expect to obtain such conditions when a magnetic field 
of 0.1~T is applied parallel to the plane of the 2DEG~\cite{Beauchamp2020}.
With the I$\pm$ ohmics fully contributing to the cooling of the 2DEG,
they dominate $G_{\text{IV}}$, Fig.~\ref{fig:G}g, therefore
$L_0 T / G_{\text{IV}} = \alpha R_{\text e} = R_{\text{I$+$\,to\,I$-$}}/4$ 
(two halves of $R_{\text{I$+$\,to\,I$-$}}$ connected in parallel).
Using the resistances measured at 1.0\,K, above $T_c$, 
$R_{\text e} = R_{\text{NT\,to\,IV}}(1.0\,{\text K}) = 27\,\Omega \  (7\,\Omega)$,
$\alpha = 0.16 \  (0.18)$ before (after) illumination, 
as measured above $T_c$. 
Then for $T_{\text{bath}} = 0.3\,$mK and $\dot Q_{\text e}^0 = 0.7\,\text{fW} \  (1.8\,\text{fW})$ Eqs.~\eqref{eq:dotQ:Tnt} and \eqref{eq:dotQ:Te} give
$T_{\text e} = 0.6\,\text{mK} \, (0.5\,\text{mK})$.
By optimising the fridge and heat exchanger performance we expect a bath temperature of 
$T_{\text{bath}} = 0.1$\,mK and an electron temperature 
in the 2DEG of $T_{\text e} = 0.5\,\text{mK} \  (0.4\,\text{mK})$.

If the superconductivity is to be tolerated, an overlayer of evaporated gold or a network of gold wirebonds
can improve the effectiveness of the large I$\pm$ contacts at least to the level of the small NT ohmic.
By scaling the effective contact resistance of the latter $R^*_{\text{NT}} = (1-\alpha) R_e^{\text{after illum.}} = 18\,\Omega$
with the perimeter we get $R^*_{\text{I}\pm} = 3\,\Omega$ for each I$+$ and I$-$. As these ohmics now dominate the smaller V$\pm$,
we estimate $G_{\text{IV}} = L_0 T / \big(R_{\text{I$+$\,to\,I$-$}}/4 + R^*_{\text{I}\pm}/2\big)$.
Assuming that the extra gold does not change the small NT ohmic significantly, we take $G_{\text{NT}}$ to be unchanged by this procedure.
This gives $R_{\text e} = 50\,\Omega \  (21\,\Omega)$, $\alpha = 0.11 \ (0.12)$ before (after) illumination leading to
$T_{\text e} = 0.6\,\text{mK}$ at $\dot Q_{\text e}^0 = 0.7\,\text{fW} \  (1.8\,\text{fW})$ and $T_{\text{bath}} = 0.1$\,mK in both cases.

\end{document}